\documentclass[superscriptaddress,aps,twocolumn]{revtex4-2}

\usepackage{graphicx}
\usepackage{dcolumn}
\usepackage{bm}
\usepackage{amsmath,amssymb,amsfonts}%
\usepackage{amsthm}%
\usepackage{mathrsfs}%
\usepackage[colorlinks=true,citecolor=blue,linkcolor=blue]{hyperref}

\begin{document}

\preprint{APS/123-QED}

\title{Towards the generation of petawatt near-infrared few-cycle light pulses via forward Raman amplification in plasma}


\author{Zhi-Yu Lei}
\affiliation{Key Laboratory for Laser Plasmas and School of Physics and Astronomy, Shanghai Jiao Tong University, Shanghai 200240, China}
\affiliation{Collaborative Innovation Centre of IFSA, Shanghai Jiao Tong University, Shanghai 200240, China}

\author{Zheng-Ming Sheng}
\affiliation{Key Laboratory for Laser Plasmas and School of Physics and Astronomy, Shanghai Jiao Tong University, Shanghai 200240, China}
\affiliation{Collaborative Innovation Centre of IFSA, Shanghai Jiao Tong University, Shanghai 200240, China}
\affiliation{Tsung-Dao Lee Institute, Shanghai Jiao Tong University, Shanghai 201210, China}
\email{zmsheng@sjtu.edu.cn}

\author{Su-Ming Weng}
\affiliation{Key Laboratory for Laser Plasmas and School of Physics and Astronomy, Shanghai Jiao Tong University, Shanghai 200240, China}
\affiliation{Collaborative Innovation Centre of IFSA, Shanghai Jiao Tong University, Shanghai 200240, China}

\author{Min Chen}
\affiliation{Key Laboratory for Laser Plasmas and School of Physics and Astronomy, Shanghai Jiao Tong University, Shanghai 200240, China}
\affiliation{Collaborative Innovation Centre of IFSA, Shanghai Jiao Tong University, Shanghai 200240, China}

\author{Jie Zhang}
\affiliation{Key Laboratory for Laser Plasmas and School of Physics and Astronomy, Shanghai Jiao Tong University, Shanghai 200240, China}
\affiliation{Collaborative Innovation Centre of IFSA, Shanghai Jiao Tong University, Shanghai 200240, China}
\affiliation{Tsung-Dao Lee Institute, Shanghai Jiao Tong University, Shanghai 201210, China}

\date{\today}

\begin{abstract}
Light amplification towards extremely high power in the infrared regime remains a significant challenge due to the lack of suitable gain media. Here we propose a new scheme to amplify a laser pulse with tunable wavelengths towards extremely high power via forward Raman amplification in plasma. Different from those previously proposed schemes based upon backward Raman or Brillouin amplification, our scheme involves a pump pulse and a seed pulse co-propagating in moderate density plasma, with the phase matching conditions for forward Raman scattering fulfilled. Due to their group velocity difference in plasma, the pump with a shorter wavelength and longer duration will chase the seed and transfer energy to the latter efficiently. Analytical models both for linear and nonlinear stages of amplification as well as particle-in-cell simulation show that by employing a $1.0\,\mathrm{\mu m}$ pump laser, a $1.8\,\mathrm{\mu m}$ seed pulse can be amplified $10^4$ times in its intensity, and then self-compressed to near-single-cycle. Our scheme shows the merits of high efficiency, high compactness, and relatively easy implementation with the co-propagating configuration, which may provide a unique route towards the petawatt few-cycle infrared laser pulses. 
\end{abstract}

\maketitle

Since the invention of laser in 1960~\cite{maiman1960stimulated}, it has become an irreplaceable tool and widely applied in numerous fields. New laser technologies such as the chirped pulse amplification~\cite{strickland1985compression} and the optical parametric chirped-pulse amplification~\cite{ross1997prospects} push the laser power to an unprecedented level of petawatt and beyond \cite{danson2015petawatt}. On the other hand, further enhancement on laser power has been challenged for years due to the limited damage thresholds both for solid-state optical materials for amplification~\cite{stuart1995laser} and for optical elements. Moreover, currently high-power lasers are limited to the wavelengths around $0.8\,\mathrm{\mu m}$ and $1.06\,\mathrm{\mu m}$, due to limitation of available gain media.

In recent years, plasma-based schemes for the amplification of high-power lasers including the Raman amplification~\cite{shvets1998superradiant,malkin1999fast,trines2011simulations,turnbull2018raman} and the Brillouin amplification~\cite{weber2013amplification,lancia2016signatures,wu2024efficient} are attracting increasing attention~\cite{riconda2023plasma}, due to that the maximum energy density the plasma can sustain is several orders of magnitude higher ($\sim 10^{17}\,\mathrm{W/cm^2}$) than that of the solid crystal ($\sim 10^{13}\,\mathrm{W/cm^2}$)~\cite{stuart1996optical}. So far both schemes all are triggered by backward scattering when two pulses are counter-propagating and colliding in the plasma. The excited electron plasma waves or ion-acoustic waves then enable the transfer of the energy from a high-energy long
pump pulse to a low-intensity short signal (or seed) pulse. However, negative effects such as kinetic instabilities~\cite{clark2003operating}, optical instabilities~\cite{langdon1975filamentation,sprangle1987relativistic}, and especially the forward scattering instability~\cite{peng2018forward}, emerge and become the major obstacles of these schemes. For example, to relieve the filamentation which deteriorates laser transverse structures, low plasma densities and low intensity pumps with a longer duration have to be taken~\cite{trines2011simulations}. These result in low energy transfer efficiency, longer plasma size, and that only seeds with wavelengths close to the pump can be amplified. Besides that, the backward scattering schemes are required to have a certain angle between two pulses to avoid head-on colliding of laser systems~\cite{vieux2017ultra}, which also increases the complexity of experiments and reduces the efficiency.

\begin{figure}[b]
\centering
\includegraphics[width=0.4\textwidth]{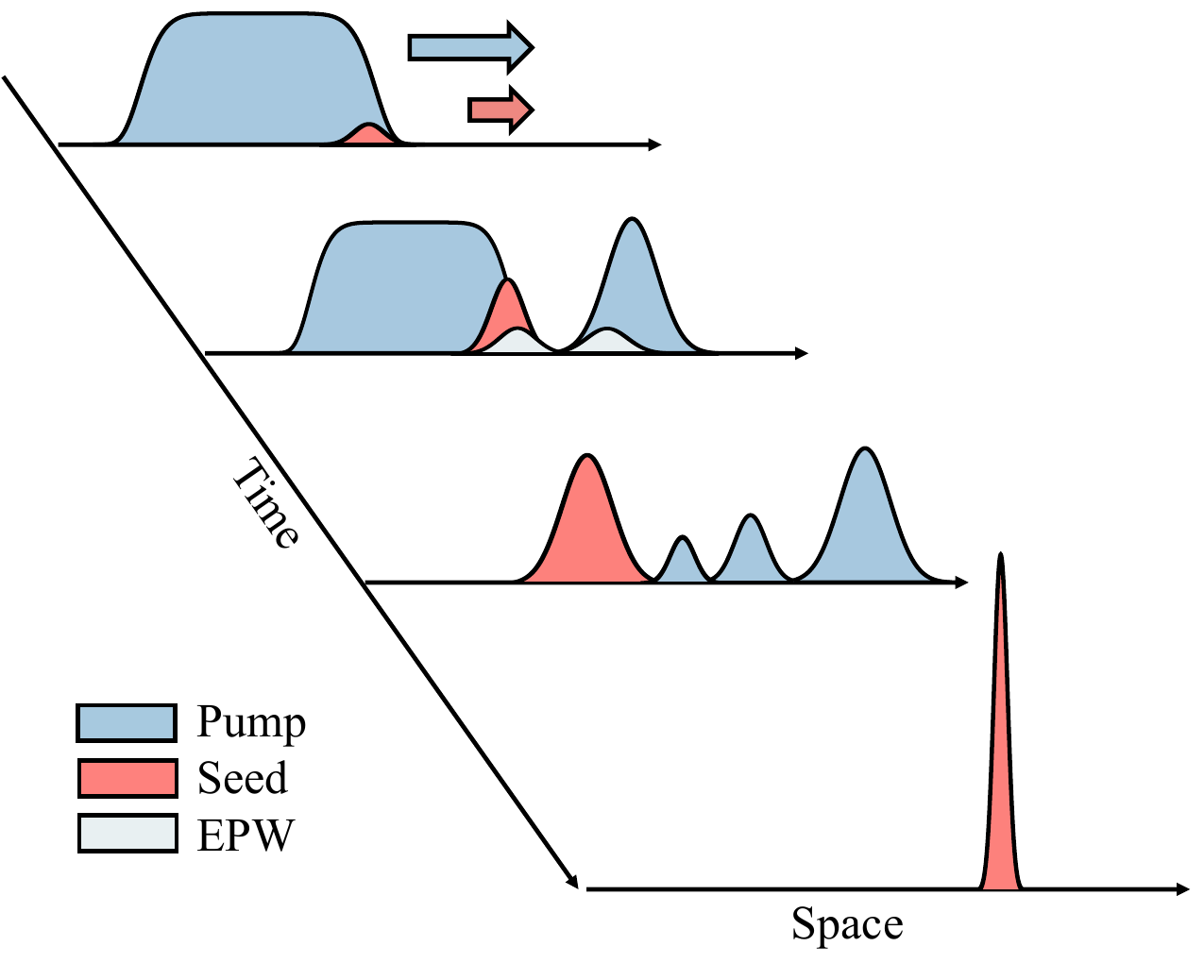}
\caption{A concept of forward Raman amplification in plasma, where the pump and seed pulses co-propagate in the forward direction and the seed pulse is located at the front of the pump pulse at the beginning. The seed pulse is well-amplified when it is overtaken by the pump. Then the seed pulse can be self-compressed to few cycles.}
\label{model}
\end{figure}

In this paper, we introduce a new plasma amplification mechanism via forward Raman scattering, i.e., forward Raman amplification (FRA), where the seed pulse and the pump pulse co-propagate in plasma as shown in Fig.~\ref{model}. Instead of trying to suppress the forward scattering instability in the backward Raman amplification scheme, we make it into the major source of light amplification.  In the FRA scheme, the pump with a shorter wavelength and longer duration chases and overtakes the seed due to their different group velocities in plasma. During their interaction, electron plasma waves are excited by the beat wave of two pulses, which continuously mediates the energy transfer from the pump to the seed. After that, the amplified seed pulse is intense enough to be self-compressed~\cite{shorokhov2003self} to nearly single-cycle via the self-phase modulation (SPM) process in plasma as it continues to propagate in plasma independently.


The analytical model of FRA is constructed below. Assume that both the pump and seed lasers co-propagate in a homogeneous plasma slab along the $+x$ direction, where the Raman scattering resonance conditions are satisfied with $\omega_0=\omega_1+\omega_2,\,\,\,\textbf{k}_0=\textbf{k}_1+\textbf{k}_2$. Here the index $0,\,1,\,2$ represent the pump, seed and electron plasma wave, respectively. The corresponding coupled three-wave equations describing the temporal and spatial evolution for the pump, seed and electron plasma wave are then given by
\begin{align}
    (\partial_t+v_0\partial_x)a_0&=\frac{ck_2}{4}\sqrt{\frac{\omega_{\mathrm{pe}}}{\omega_0}}a_1a_2,\\
    (\partial_t+ v_1\partial_x)a_1&=-\frac{ck_2}{4}\sqrt{\frac{\omega_{\mathrm{pe}}\omega_0}{\omega_1^2}}a_0a_2^*\label{3w},\\
    (\partial_t+v_2\partial_x)a_2&=-\frac{ck_2}{4}\frac{\omega_{\mathrm{pe}}}{\omega_2}\sqrt{\frac{\omega_{\mathrm{pe}}}{\omega_0}}a_0a_1^*,
\end{align}
where $a_{0,1,2}$ are the envelope amplitudes of three waves, and $v_{0,1,2}=\partial\omega_{0,1,2}/\partial k_{0,1,2}$ are the group velocities. For simplicity, we consider perfect phase matching condition and cold plasma, so that $v_2\approx 0$ and $\omega_2\approx\omega_{\mathrm{pe}}$.

\begin{figure}[t]
\centering
\includegraphics[width=0.48\textwidth]{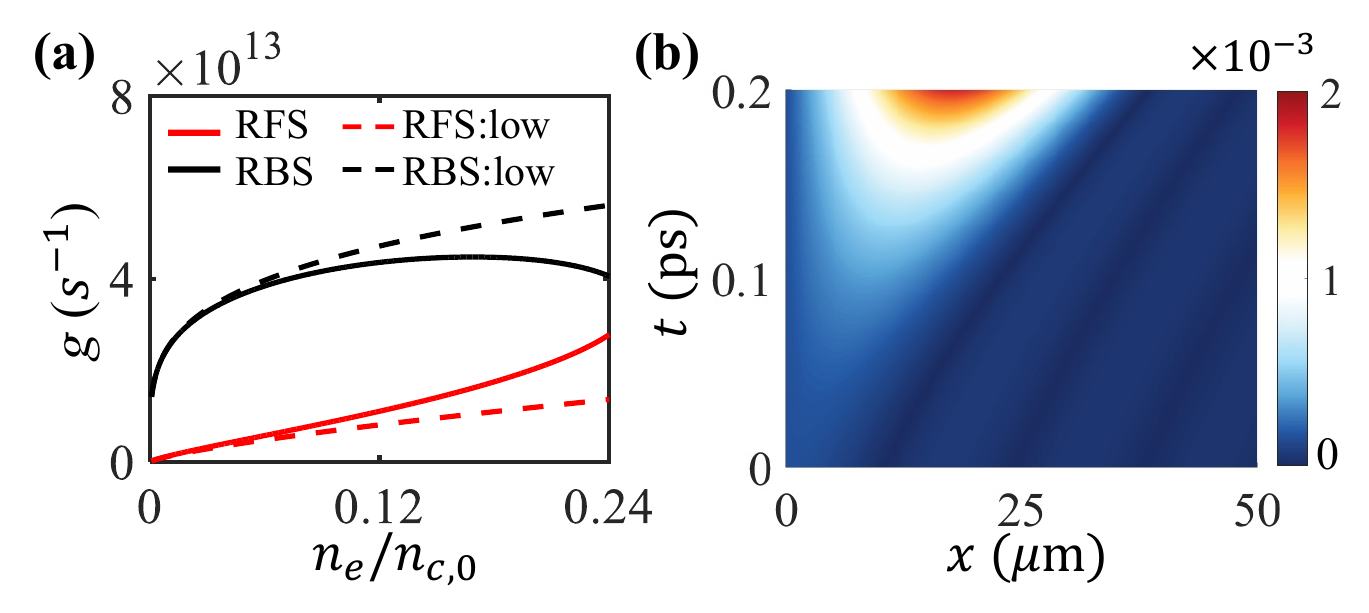}
\caption{Analytical results of the forward Raman amplification in the linear stage. (a) The growth rates $g$ in forward (red) and backward (black) cases with different plasma density. The dashed lines are the low-density approximation results. (b) The temporal and spatial envelope evolution of the seed pulse $a_1(\zeta,\tau)$, and transformed into the real space $(x,t)$. The parameters used for both (a) and (b) are $a_{0,0}=0.0854,\,\,\,a_{1,0}=0.0015,\,\,\,\lambda_0=1.0\,\mathrm{\mu m},\,\,\,\lambda_1=1.8\mathrm{\mu m}$, and $n_e=0.2\,n_{c,0}$.}
\label{linear_plot}
\end{figure}

Let us first consider the linear stage of Raman amplification. In this case, the pump depletion is negligible~\cite{malkin2000detuned}, and $a_0=a_{0,0}$ is a constant. By introducing the new coordinates $\tau=x/v_1,\,\,\,\zeta=t-x/v_1$, and assuming a fixed boundary condition of the seed $a_1(\zeta,0)=a_{1,0}$, the three-wave equation can be solved analytically via the Laplace transformations, the amplitude of seed is found to be  $a_1(\zeta,\tau)=a_{1,0}I_0(2g\sqrt{\zeta\tau})$ (see Supplementary). Here $I_0$ is the modified Bessel function, and $g$ is the growth rate of Raman linear amplification given by
\begin{equation}
    g=a_0ck_2\sqrt{\omega_{\mathrm{pe}}/(\omega_0-\omega_{\mathrm{pe}})}/4,
    \label{g_rate}
\end{equation}
where $k_2=(\sqrt{\omega_0^2-\omega_{\mathrm{pe}}^2}\mp\sqrt{\omega_0^2-2\omega_0\omega_{\mathrm{pe}}})/c$ is determined by the linear dispersion relations of three waves for the Raman forward (backward) scattering. At a low plasma density, Eq.~\eqref{g_rate} reduces to the well-known expressions $g_{\mathrm{RFS}}\approx a_0\omega_{\mathrm{pe}}\sqrt{\omega_{\mathrm{pe}}/\omega_0}/4$ and $g_{\mathrm{RBS}}\approx a_0\sqrt{\omega_{\mathrm{pe}}\omega_0}/2$, respectively, for the undamped Raman instability growth rates for forward scattering~\cite{decker1996spatial,schroeder2003raman} and backward scattering~\cite{forslund1975theory}. In Fig.~\ref{linear_plot}(a), we present the Raman linear growth rates as a function of plasma density both for forward and backward cases according to Eq.~\eqref{g_rate}. It is found that $g_{\mathrm{RFS}}$ increases continuously with the plasma density and the difference between backward and forward amplification rates shrinks with increasing plasma density. The FRA features in the linear stage given by the analytical solution of seed pulse $a_1$ is shown in Fig.~\ref{linear_plot}(b). The intensity of the seed pulse is amplified exponentially as co-propagating with the pump pulse in plasma. As a whole, since the interaction time for the forward case is longer due to the smaller relative velocity between the pump and seed pulses, it is expected that the final FRA amplitude shall be comparable to that in the backward case.

\begin{figure}[t]
\centering
\includegraphics[width=0.4\textwidth]{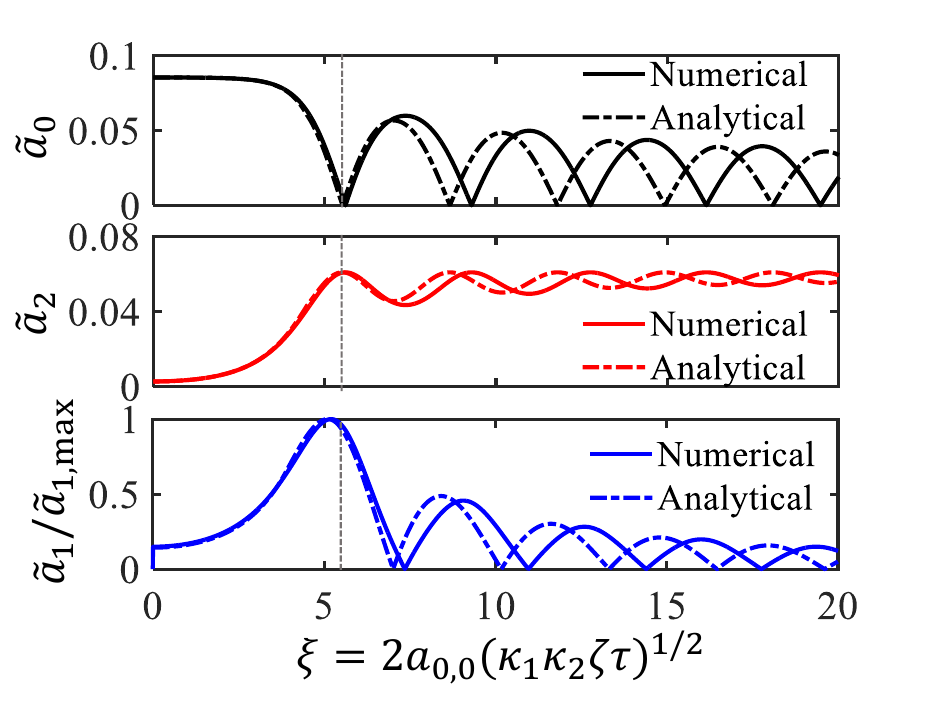}
\caption{Evolution of three waves in the nonlinear stage of the forward Raman amplification, where the solid lines are the numerical results by solving sine-
Gordon equation numerically, and the dotted-dash lines are the analytical solutions under the Bessel approximation. The parameters are the same as in Fig.~\ref{linear_plot} and, in addition, the seed and pump pulses have a duration of $20T_1$ and $40T_1$, respectively.}
\label{nonlinear_plot}
\end{figure}

For the nonlinear stage of FRA, we choose the new space coordinate $\tau=ck_2\sqrt{\omega_{\mathrm{pe}}/\omega_0}/4\cdot x/v_1$  and the co-moving time coordinate $\zeta=ck_2\sqrt{\omega_{\mathrm{pe}}/\omega_0}/4\cdot(t-x/v_1)$. In the nonlinear stage, strong pump depletion is expected and the pump changes rapidly with  $\zeta$ and slowly with $\tau$~\cite{lehmann2013pulse}. Thus, the $\partial_\tau$ term can be ignored compared to $\partial_\zeta$ term for the pump. As a result, the three-wave equations become $\partial_{\zeta}a_0=\kappa_0a_1a_2,\,\,\,\partial_{\tau}a_1=\kappa_1a_0a_2^*,\,\,\,\partial_{\zeta}=\kappa_2a_0a_1^*$. Here $\kappa_0=1/(1-v_0/v_1),\,\,\,\kappa_1=-\omega_0/\omega_1,\,\,\,\kappa_2=-1$. In order to solve above equations analytically, we introduce the self-similar function $\xi=2a_{0,0}\sqrt{\kappa_1\kappa_2\zeta\tau}$. Based on that, the three-wave equations then can be solved as $ \tilde{a}_0=a_{0,0}\cosh(\frac{\Tilde{u}}{2}),\,\,\,\tilde{a}_1=a_{0,0}^2\kappa_1\sqrt{\frac{\kappa_2}{\kappa_0}}\cdot\xi^{-1}\frac{d\tilde{u}}{d\xi},\,\,\,\tilde{a}_2=a_{0,0}\sqrt{\frac{\kappa_2}{\kappa_0}}\sinh(\frac{\Tilde{u}}{2})$(see Supplementary), where $\Tilde{\nu}(\xi)=i \Tilde{u}(\xi)$ is the solution of the sine-Gordon equation $\Tilde{\nu}_{\xi \xi}+\Tilde{\nu}_{\xi}/\xi=\sin(\Tilde{\nu})$, and can be approximated by $\Tilde{\nu}\approx 2\sqrt{\kappa_0\kappa_2}|a_{1,0}|\delta(\zeta)J_0(i\xi)$ for $\xi<\xi_0$, and $\Tilde{\nu}\approx \pi(1-1/2J_0(\xi_0))J_0(\xi)$ for $\xi>\xi_0$. Here $\delta(\zeta)$ is the initial seed pulse profile with a narrow duration, $J_0$ is the zeroth Bessel function of the first kind, and $\xi_0$ is the point where $\tilde{\nu}(\xi_0)=\pi/2$. Figure~\ref{nonlinear_plot} illustrates the basic features for the evolution of the three waves in the nonlinear stage of FRA. It is found that the seed pulse grows remarkably with the pump amplitude decreasing. Once the pump is depleted to zero and the seed reaches to its maximum, the energy is converted reversely from the seed to the pump until the pump amplitude reaches a peak. Such a periodic energy conversion process continues, causing a wave train named ``$\pi$ pulse" in both two lasers, which are almost the same pattern as in the backward Raman amplification~\cite{malkin1999fast}. By the expansion of the Bessel function $J_0$ for small $\xi$, one can further derive the approximate solution of the seed pulse intensity at the early nonlinear stage
\begin{equation}
a_1^2\approx a_{0,0}^4 a_{1,0}^{2}(\kappa_1\kappa_2)^2\delta(\zeta)^2\tau^2.\label{nonlinear_app}
\end{equation}
This analytical result suggests that the intensity of the amplified seed pulse $I_1$ is proportional to the square of the propagation distance $\tau$, which is a typical feature for superradiant amplification~\cite{ersfeld2005superradiant}. Equation \ref{nonlinear_app} also illustrates the parameter dependence of the seed on the pump intensity with $I_1\propto a_{0,0}^4$, and the background plasma density $I_1\sim n_e$. Such quantitative scaling relations can provide guidance for numerical simulations and even future experimental investigations.

Above analysis well illustrates the effectiveness of FRA. However, the seed pulse duration after FRA is limited to the scale of $g^{-1}$ (or $\sim100\,\mathrm{fs}$), which may cause an enlargement of the seed duration if it is an initially narrow one. Fortunately, self-compression can be triggered by the amplified seed as it continuously propagates in the moderate density background plasma. The time scale for the full self-compression is estimated to be $\tau_{\mathrm{sfc}}=cD/2\omega v_g$, where $D=\pi\delta T_0^2(n_c/n)[(1-n/n_c)/(\delta-1)]^{3/2}/2$. It decreases as the plasma density increases. However, a higher density also results in a stronger filamentation instability. In principle, one can isolate the filamentation from the amplification as the required pump duration is short. Nevertheless, the self-compression process could be deteriorated by the filamentation instability since the growth rate of filamentation $\gamma_{\mathrm{fil}}=(\omega_{\mathrm{pe}}^2/8\omega)a^2$~\cite{langdon1975filamentation} may be comparable to the inverse self-compression time $1/\tau_{\mathrm{sfc}}$. Therefore, the laser and plasma parameters for FRA should comply with the following constraint to achieve ideal performance: (i) The linear amplification should be efficient enough so that $\tau_{\mathrm{RFS}}<\tau_{\mathrm{window}}$. Here $\tau_{\mathrm{RFS}}=1/g_{\mathrm{RFS}}$ is the time scale of the linear stage in FRA, $\tau_{\mathrm{window}}=c\tau_{\mathrm{pump}}/(v_{g,\mathrm{pump}}-v_{g,\mathrm{seed}})$ is the interaction time window between the pump (with duration $\tau_{\mathrm{pump}}$) and the seed; (ii) In order to avoid the considerable development of the filamentation instability during the amplification, the interaction time should be well controlled so that $\tau_{\mathrm{window}}<\tau_{\mathrm{fil}}$; (iii) One needs to make sure the filamentation is not significant before the self-compression of the amplified seed is completed $\tau_{\mathrm{sfc}}<\tau_{\mathrm{fil}}$.

\begin{figure}[b]
\centering
\includegraphics[width=0.45\textwidth]{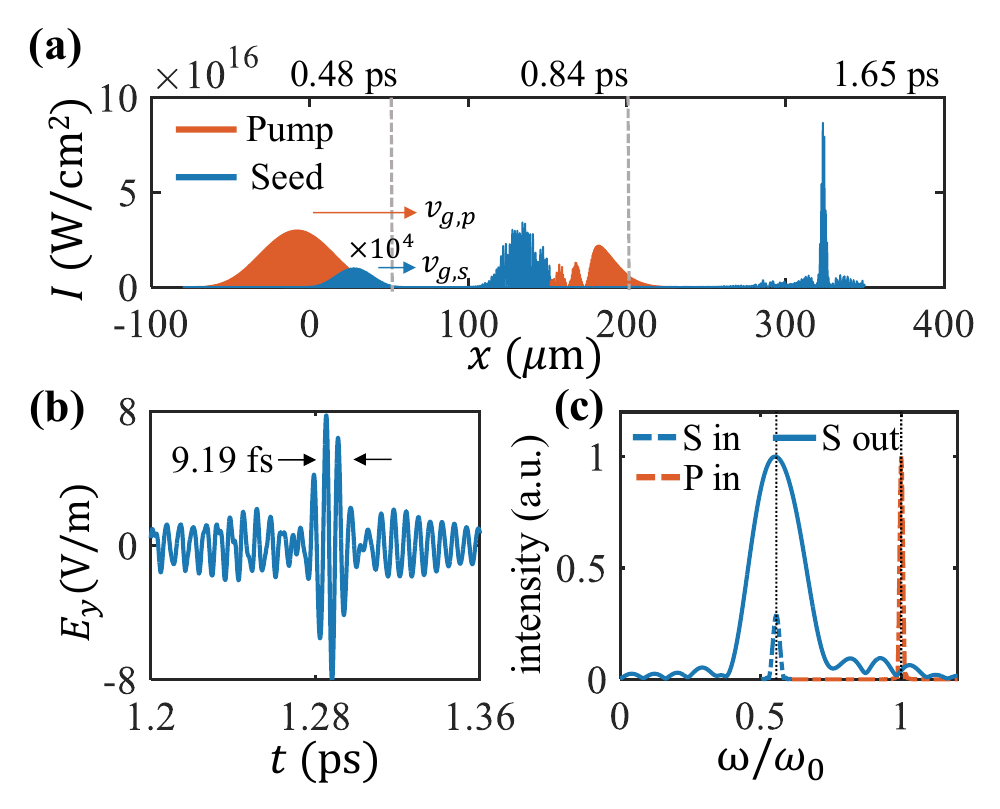}
\caption{Seed pulse amplification and compression in FRA scheme. (a) The spatial distributions of the seed (blue) and pump (red) intensity at different time (seed intensity $\times 10^4$ magnified to make it visible). The boundaries of the uniform plasma are marked by the gray dashed lines. (b) The output seed pulse in vacuum as a function of time. (c) Frequency spectra of the both incident pump (P) and seed (S) laser pulses and the output seed pulse. Here $\omega_0$ is the central frequency of the pump, and the dotted lines are the analytical central frequency ratio based on their wavelengths.}
\label{1D_long}
\end{figure}
We start by carrying out one-dimensional (1D) PIC simulation in order to demonstrate the FRA scheme. Figure \ref{1D_long} presents the 1D PIC simulation results obtained with EPOCH~\cite{arber2015contemporary}, showing the amplification of a seed pulse at $1.8\,\mathrm{\mu m}$ by a pump pulse at $1.0\,\mathrm{\mu m}$. The seed pulse is initially with a duration of $120\,\mathrm{fs}$ and the peak intensity of $1.0\times 10^{12}\,\mathrm{W/cm^2}$. The pump pulse is initially with a duration of $240\,\mathrm{fs}$ and the peak intensity of $3.0\times 10^{16}\,\mathrm{W/cm^2}$. With their fronts overlapped as shown in Fig.~\ref{1D_long}(a), they are co-incident into a uniform plasma slab with the density  $2.2\times 10^{20}\,\mathrm{cm^{-3}}$ (i.e., $0.2\,n_{c,\mathrm{pump}}$ or $0.64\,n_{c,\mathrm{seed}}$) to match the 3-wave coupling condition of forward Raman scattering. The background plasma composed of mobile electrons and protons is cold and uniformly distributed within $50\,\mathrm{\mu m}$ to $200\,\mathrm{\mu m}$ in the simulation box. During their co-propagation in plasma, the seed pulse is gradually overtaken by the pump for its longer wavelength, and thus a smaller group velocity than the pump which is given by $v_{g,\mathrm{seed}}=c\sqrt{1-\omega_{\mathrm{pe}}^2/\omega_{\mathrm{seed}}^2}$. The seed then gets amplification dynamically. When the pump has passed through the seed completely, the amplified seed pulse continuously propagates independently in the plasma and trigger self-compression, resulting in significant compression in its duration and further dramatic increase in its peak intensity. The final output intensity of the seed pulse is amplified about $10^{5}$ times from $1.0\times 10^{12}\,\mathrm{W/cm^2}$ to $8.9\times 10^{16}\,\mathrm{W/cm^2}$, and the duration is compressed from $120\,\mathrm{fs}$ to near-single-cycle to $9.19\,\mathrm{fs}$. One may notice that the depleted pump appears in the form of a wave train, which is the same as that predicted by our theoretical model shown in above, indicating that the amplification has reached the nonlinear regime. In the simulation, a virtual probe is placed at the right side vacuum area to detect the electric field of the amplified seed pulse as a function of time (Fig.~\ref{1D_long}(b)). By evaluating the frequency spectra of both the input and output pulses, as shown in Fig.~\ref{1D_long}(c), we further confirm that the seed pulse has been amplified and compressed efficiently by forward Raman scattering combined with self-compression in the plasma. For another case that the initial seed pulse has an extreme short duration, the results are discussed in the Supplementary Materials.

\begin{figure}[t]
\centering
\includegraphics[width=0.48\textwidth]{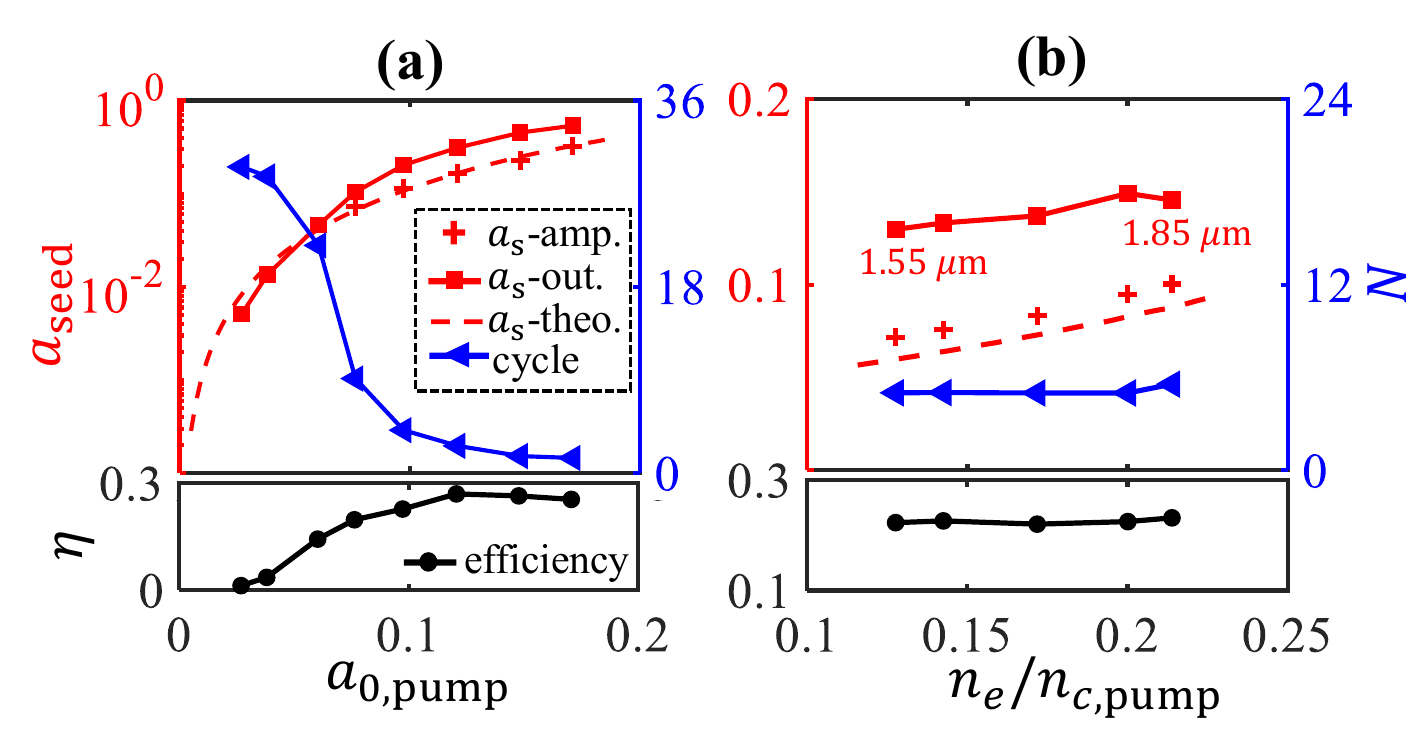}
\caption{Effects of the laser and plasma parameters on the output seed pulse. (a) The change of the seed intensity $a_\mathrm{seed}$ (red), optical cycle number $N$ (blue), and energy conversion efficiency $\eta$ (black) with the pump intensity. Other parameters are the same as those in Fig.\ref{1D_long}; (b) The change of the output seed parameters with the plasma density with fixed $a_{0,\mathrm{pump}}=0.0854$ and $a_{0,\mathrm{seed}}=0.0015$.}
\label{parameters}
\end{figure}

We then investigate the effects of the pump intensity and plasma density on light amplification and compression by FRA. For a given plasma density with $0.2n_{c,\mathrm{pump}}$, as the pump intensity increases, Fig. \ref{parameters}(a) shows that the output seed pulse intensity has a quadratic growth to $1.0\times 10^{17}\,\mathrm{W/cm^2}$ and the duration decreases to nearly single-cycle. The energy transfer efficiency grows as well and saturates at around $30\%$. 
Meanwhile, it is noticed that as the pump amplitude increases beyond $a_{0,\mathrm{pump}}=0.06$, the final output seed intensity (red square line) starts to be much higher than the theoretical prediction (red dashed line) given by Eq.~\eqref{nonlinear_app}, and the seed duration drops instantly as well. This is due to the fact that the self-compression will be triggered when the seed is amplified to an enough large value, causing a further increase in its strength and decrease of its duration. Because of this, we now only consider the seed intensity when the pump has just overtaken the seed pulse before significant self-compression.  In this case, the intensity amplification is mainly contributed by FRA, while the compression is not remarkable at that time. As shown in Fig.~\ref{parameters}(a), the intensity values obtained at this stage (red pulse sign) from simulation are well fitted in the theoretical model. This verifies our theoretical analysis (Eq.~\eqref{nonlinear_app}) that one shall have $a_{\mathrm{seed}}\sim a_{0,\mathrm{pump}}^2$ for pure FRA.

In Fig.~\ref{parameters}(b), we illustrate the effect of the background plasma density on the seed pulse amplification and compression. As the plasma density $n_e$ increases with a fixed pump pulse wavelength, the corresponding amplified seed wavelength shall grow as well to match the 3-wave coupling condition, given by $\lambda_{\mathrm{seed}}=\lambda_{\mathrm{pump}}/(1-\sqrt{n_e/n_{c,\mathrm{pump}}})$. It is found that the final output seed intensity increases slowly, while the duration and efficiency are almost unchanged. Due to the self-compression, the final output seed intensity is larger than that by pure FRA. The latter is in good agreement with the theoretical model given by Eq.~\eqref{nonlinear_app}. It is noted that for a longer seed wavelength, the interaction time and the required plasma length become shorter, while the output intensity and duration are slightly affected. Above results suggest that FRA and subsequent self-compression work effectively for the seed pulse with different wavelengths and pump pulse intensity as long as a matched plasma density is applied.

\begin{figure}[t]
\centering
\includegraphics[width=0.48\textwidth]{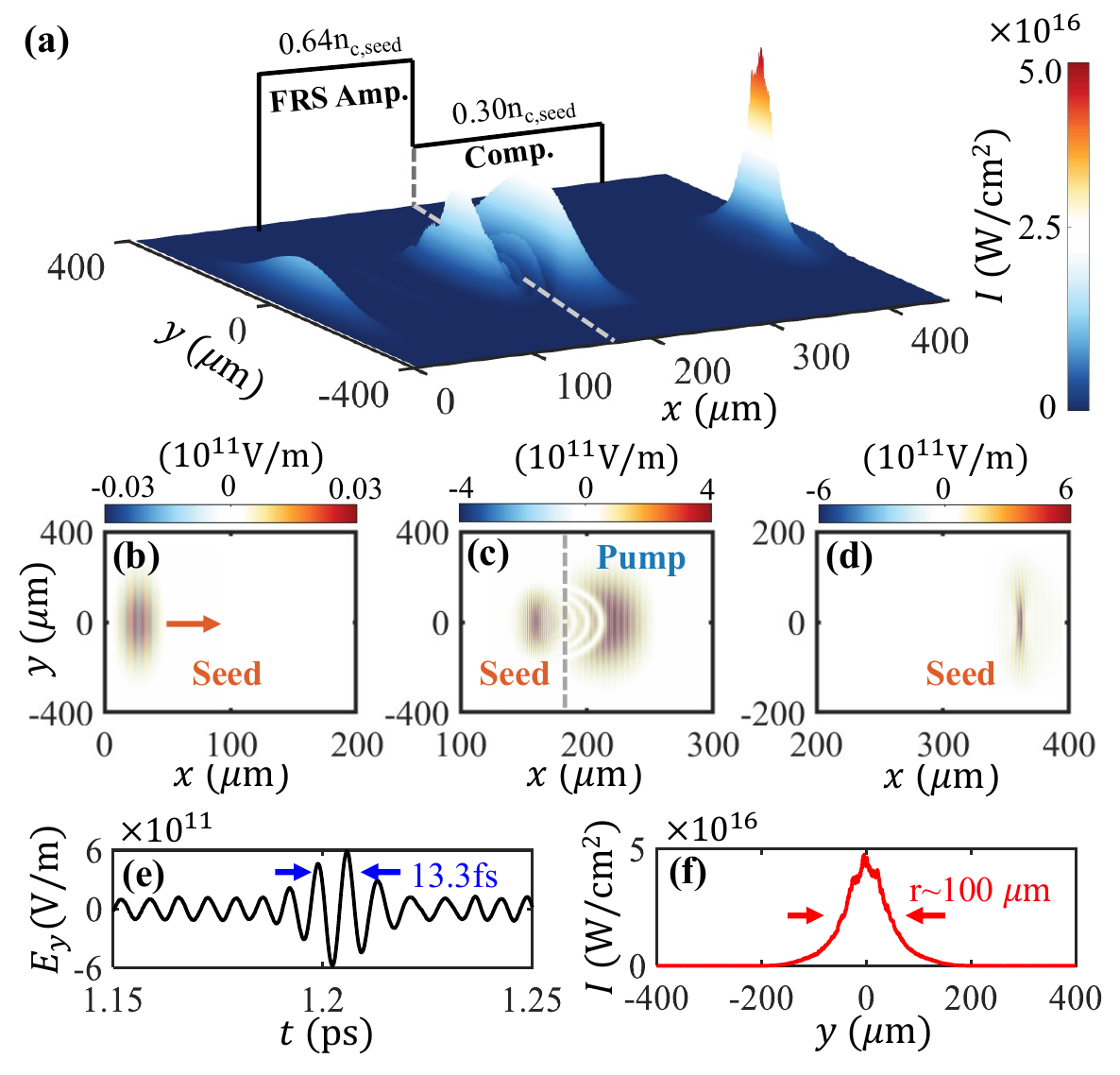}
\caption{(a) The pulse intensity obtained by 2D PIC via forward Raman amplification and self-compression. (b)$\sim$(d) The spatial distributions of electric fields of the seed and pump at different stages, where the gray dashed line in (c) is the density boundary between amplification and compression. (f) The final output seed electric fields detected in vacuum as a function of time. (g) The transverse intensity distributions of the output seed pulse.}
\label{2D}
\end{figure}

A case from 2D PIC simulation is also studied. As shown in Fig.~\ref{2D}, a seed pulse of $1.8\,\mathrm{\mu m}$ is adopted with initial intensity $1.0\times 10^{12}\,\mathrm{W/cm^2}$, duration $90\,\mathrm{fs}$, and transverse spot radius $200\,\mathrm{\mu m}$ with the Gaussian distribution. The pump wavelength is $1.0\,\mathrm{\mu m}$, with initial intensity $2.5\times 10^{16}\,\mathrm{W/cm^2}$, duration $180\,\mathrm{fs}$ and the same spot size as the seed. The plasma density distribution is divided into two parts. One is $0.64\,n_{c,\mathrm{seed}}$ for FRA and the other one is $0.3\,n_{c,\mathrm{seed}}$ for self-compression. Here we employ a lower plasma density for seed pulse self-compression to reduce the filamentation instability. Such a plasma density distribution can be achieved by the two-chamber gas target in experiments~\cite{drobniak2023two}. With the setup above, it is found that the seed pulse is amplified to $4.9\times 10^{16}\,\mathrm{W/cm^2}$ and compressed to $13.3\,\mathrm{fs}$ with the output spot size of $100\,\mathrm{\mu m}$. 
Figures~\ref{2D}(b)$\sim$\ref{2D}(d) clearly illustrate the evolution of the seed pulse. The seed is first amplified by the pump as the latter gradually overtakes the seed, and continuously converts the energy to the seed through the electron plasma wave generated by forward Raman scattering. Then the amplified seed triggers the self-compression. The above scenario is the same as shown in 1D PIC results. The output seed spot size is half of the initial one due to the fact that the seed intensity in the outer region is too small to be amplified effectively. Such phenomenon is common and has been observed in previous backward Raman amplification experiments~\cite{ren2007new}. By further analysing the longitudinal and transverse distributions of the output seed pulse, Fig.~\ref{2D}(f) shows that it is well compressed. Meanwhile, the filamentation instability is controlled effectively as no significant splitting or filaments are observed in the transverse distribution of the final seed pulse, as shown in Fig.~\ref{2D}(g).

The advantages of FRA scheme are summarized as follows. Due to that the co-propagating configuration has smaller relative speed between two pulses, the amplification can achieve more sufficient interaction between the pump and seed even with limited pulse duration. This allows shorter amplification time and plasma size, which shall suppress the instabilities growing significantly, and reduce the long-scale plasma inhomogeneity effects. Besides that, kinetic effects like Landau damping or wavebreak are less concerned due to the fact that the phase velocity of the plasma wave in the forward scattering is larger than the backward case, so that the particles are less likely to be captured by waves.

In conclusion, we have proposed a new scheme to efficiently amplify a seed laser pulse with tunable wavelength via forward Raman amplification, where a pump laser pulse co-propagating with the seed pulse is utilized. Energy transfer from the pump to the seed is realized via three-wave resonant coupling. A corresponding theoretical model for linear and nonlinear amplification is constructed and verified, including an analytical scaling relation of the amplified seed pulses with the pump and the plasma density. By 1D and 2D PIC simulations, we demonstrate that an initially extreme weak seed pulse with longer wavelength can be amplified $10^{4}\sim 10^{5}$ times to well above the intensity of $10^{16} {\rm W/cm^2}$ within a short distance. Moreover, the amplified seed pulse can be self-compressed to the nearly single-cycle in plasma without significant development of instabilities. 
The FRA scheme with relatively easy implementation in experiments works for various laser and plasma parameters, and is promising to achieve longer wavelength amplification via cascading the amplified seed as the new pump pulse. It provides unique advantages of generation petawatt few-cycle near-infrared laser pulses with tunable wavelengths in particular, by simply tuning the plasma density. This may offer a unique opportunity for ultrafast science and high field applications.

This work is supported by the the Strategic Priority Research Program of Chinese Academy of Sciences (Grant Nos. XDA25050100 and XDA25010100), National Natural Science Foundation of China (Grant Nos. 12135009, 11991074, and 12005287). Numerical simulations were performed on Computer $\pi$2.0 in the Center for High Performance Computing at Shanghai Jiao Tong University.


\bibliography{apssamp}

\end{document}